\documentclass[pdflatex,sn-mathphys-num]{sn-jnl}%

\usepackage{graphicx}%
\usepackage{multirow}%
\usepackage{amsmath,amssymb,amsfonts}%
\usepackage{amsthm}%
\usepackage{mathrsfs}%
\usepackage[title]{appendix}%
\usepackage{textcomp}%
\usepackage{manyfoot}%
\usepackage{booktabs}%
\usepackage{algorithm}%
\usepackage{algorithmicx}%
\usepackage{algpseudocode}%
\usepackage{listings}%
\usepackage{bm}
\usepackage{array}
\usepackage{tabularx}
 \usepackage{multirow}
 \usepackage[table,xcdraw]{xcolor}
\usepackage[export]{adjustbox}
\usepackage{float}
\restylefloat{table}
\usepackage{booktabs}
\usepackage{placeins}  
\usepackage{mathtools}

\theoremstyle{thmstyleone}%
\theoremstyle{thmstyletwo}%

\theoremstyle{thmstylethree}%

\begin{document}

\title[Article Title]{Unified Beta Regression Model with Random Effects for the Analysis of Sensory Attributes}

\author*[1]{\fnm{João César} \sur{Reis Alves}}\email{joaocesar@usp.br}
\equalcont{These authors contributed equally to this work.}

\author[2]{\fnm{Gabriel} \sur{Rodrigues Palma}}\email{gabriel.palma.2022@mumail.ie}
\equalcont{These authors contributed equally to this work.}

\author[1]{\fnm{Idemauro} \sur{Antonio Rodrigues de Lara}}\email{idemauro@usp.br}

\affil*[1]{\orgdiv{Department of Exact Sciences}, \orgname{"Luiz de Queiroz" College of Agriculture, University of São Paulo}, \orgaddress{ \city{Piracicaba},  \state{São Paulo}, \country{Brazil}}}

\affil[2]{\orgdiv{Hamilton Institute}, \orgname{Maynooth University}, \orgaddress{ \city{Maynooth}, \state{County Kildare}, \country{Ireland}}}


\abstract{Studies involving sensory analysis are essential for evaluating and measuring the characteristics of food and beverages, including consumer acceptance of samples. For various products, the experimental designs are generally incomplete block designs, with sensory attributes assessed using hedonic scales, ratings, or scores. Statistical methods such as generalized logits are commonly used to analyze these data but face limitations, including convergence issues due to superparameterization. Furthermore, sensory attributes are traditionally analyzed separately, increasing the complexity of the process and complicating the interpretation of results. This study proposes a unified beta regression model with random effects for simultaneously analyzing multiple sensory attributes, whose scores were converted to the (0,1) interval. Simulation studies demonstrated overall agreement rates greater than 82\% for the unified model compared to models fitted separately for each attribute. As a motivational example, the unified model was applied to a real dataset in which 98 potential consumers evaluated eight grape juice formulations for each sensory attribute: colour, flavour, aroma, acidity, and sweetness. The unified model identified the same top-rated formulations as the separately fitted models, characterized by a higher proportion of juice relative to sugar. The results underscore the ability of the unified model to simplify the analytical process without compromising accuracy, offering an efficient and insightful approach to sensory studies.}


\maketitle

\section{Introduction}\label{sec1}

The evaluation of the organoleptic profile of food, cosmetic, chemical, and textile products is often carried out through a procedure known as sensory analysis \cite{zeng2008overview} \citep{piana2004sensory}. In this context, sensory attributes or characteristics such as sweetness, colour, flavour, aroma, overall impression, and body are assessed based on a previously defined response scale. This scale, structured in ordinal categories, establishes a hierarchy that reflects different levels of preference and is known as the hedonic scale \citep{bergara2002hedonic}. Among the various hedonic scales available, the five-point ordinal scale, which ranges from least to most favourable, has been widely used due to its practicality, especially in studies involving consumers with limited familiarity with sensory tests or in settings that require a more streamlined data collection process \citep{atanasova2018hedonic}. Sensory evaluation methodologies aim to measure responses with minimal bias \citep{lopetcharat201112}. However, limitations such as panellist fatigue, restricted sample capacity, and spatial constraints often require using balanced incomplete block (BIB) designs to control experimental errors and ensure reliable results \citep{ball1997incomplete}. In many studies, BIB designs are employed with hedonic scales for product evaluation. 

Although hedonic scales are widely used, their ordinal nature raises important considerations for statistical analysis. Normal linear models, such as regression and ANOVA, are commonly applied to these data, often treating categorical ratings as continuous points \citep{christensen2013analysis}. However, some authors caution against the use of ANOVA in this context \citep{o2017sensory, villanueva2000performance}, as this approach, while suitable in some instances, can lead to imprecise parameter estimates and reduced test power, especially when there are few categories or extreme responses \citep{smithson2006better}. For instance, \citep{song2023sensory} analyzed the sensory quality of juices using two-way and three-way ANOVA with Tukey's HSD to identify drivers of consumer preferences. Similarly, \citep{abeywickrema2024evaluating} examined differences in olfactory perception between food groups using ANOVA and Spearman's correlation, highlighting the critical role of selecting appropriate statistical methods in sensory analysis.

The application of cumulative logit models is considered more appropriate for these trials, as these models account for the ordered and categorical nature of the data \citep{oehm2022identifying, gutierrez2015ordinal}. Examples of this approach are presented in the works of \citep{gadrich2022ordinal}, \citep{christensen2013analysis}, and \citep{kallas2016comparison}, while a more in-depth description of these models is available in \citep{agresti2012categorical, balbino2023analise}. Despite their advantages, these models have the limitation of being superparameterized, which often requires data grouping to make their application viable. Furthermore, statistical models tend to be applied separately to each sensory attribute evaluated in sensory analysis studies. This practice results in a certain computational complexity due to the fragmentation of the analysis, making it difficult to obtain an integrated view of the sensory attributes. As a consequence, it becomes challenging to understand consumer preferences comprehensively.

This research presents a unified mixed beta regression model designed to analyze the evaluated sensory attributes jointly. The beta distribution belongs to the exponential family, is highly flexible, can assume various shapes, and accounts for heteroscedasticity, non-normality, and skewness in the data \citep{johnson1995continuous}. The beta regression model, a specific case of generalized linear mixed models (GLMMs), was introduced by \citep{ferrari2004beta} and is frequently utilized when the dependent variable explains the behaviour of rates and proportions that take values in the interval 
$(0,1)$ \citep{qasim2021some}. This model is applied across various fields, including engineering, medical sciences, physical sciences, social sciences, environmental studies, and business \citep{abonazel2022dawoud}.

Furthermore, beta regression models minimize issues of superparameterization and do not require data grouping \citep{junaid2024modified}. Therefore, this approach is expected to be useful for researchers in the field of sensory analysis, providing a technique that supports more assertive decision-making. The paper unfolds as follows. Section \ref{method} is divided into two subsections. Subsection \ref{method1} reviews the regression model and its extension to incorporate random effects. Subsection \ref{method2} specifies the beta regression model used for the individual analysis of the sensory attributes. In Subsection \ref{unifi}, the proposed unified beta model is introduced, along with a discussion of the maximum likelihood inference procedure. Section \ref{simulation} is devoted to a simulation study designed to evaluate the performance of the proposed model. Section \ref{aplication} presents an application of the unified beta model to sensory data from eight grape juice formulations under a BIB design. Concluding remarks and suggestions for future research are provided in Section \ref{conclusion}.

\section{Method} \label{method} \subsection{Review on beta regression model} \label{method1}

Let $Y \sim \text{Beta}~(\mu, \phi)$, with the probability density function (p.d.f.): \begin{equation} f(y, \mu , \phi ) = \frac{\Gamma (\phi )}{\Gamma (\mu \phi )\Gamma ((1 - \mu )\phi )} y^{\mu \phi -1} (1 - y)^{(1 - \mu )\phi -1} \mathbb{I}_{(0,1)}(y) \end{equation} where $0 < \mu < 1$ and $\phi \in \mathbb{R}^+$ is the precision parameter. This parameterization, proposed by \citep{ferrari2004beta}, implies that for a fixed mean $\mu$, larger values of $\phi$ correspond to lower variance in the dependent variable. The beta distribution is part of the exponential family, though not in canonical form \citep{brown1986fundamentals}.

Now, let $y_1, \dots, y_n$ be $n$ independent random variables such that $y_i \sim \text{Beta}(\mu \phi, (1 - \mu)\phi)$. The definition of a beta regression model requires a transformation of the mean $\mu_i$ of $y_i$, $i = 1, \dots, n$, that maps the interval $(0,1)$ onto the real line. A commonly used and practical link function is the logit link, primarily due to its ease of interpretation in practical applications. Thus, it is assumed that \begin{equation}\label{mod2} \ln\left( \frac{\mu_{i}}{1 - \mu_{i}} \right) = \textbf{x}_i^{\top} \bm{\beta}, \end{equation} where $\textbf{x}_i$ is a vector of known covariates for the $i$-th subject, and $\bm{\beta}$ denotes a vector of regression coefficients \citep{figueroa2013mixed}. In this framework, the precision parameter $\phi$ is typically treated as a nuisance parameter, i.e., it is constant across all observations.

In many situations, repeated observations for the same individuals may occur. This necessitates an extension of the beta regression model (\ref{mod2}) through the inclusion of random effects. Such an extension is particularly important in longitudinal data analysis, where serial dependence between observations is common \citep{hunger2012longitudinal}. It is also relevant in experimental designs where blocks are treated as a random structure within the model.

The random-effects beta regression model is an extension of Generalized Linear Mixed Models (GLMMs) \citep{smithson2006better}, in which the linear predictor integrates both fixed and random effects, accommodating dependencies among observations \citep{mccullagh2019generalized}. This approach is particularly valuable for repeated non-Gaussian measurements, given the flexibility of the beta distribution in modeling dependent variables with characteristics such as asymmetry and bimodality \citep{verkuilen2012mixed}.

In this scenario, consider $\bm{y}_i = (y_{i1}, \dots, y_{in_i})^\top$ as an individual response profile for a sample unit $i$, in which each component $y_{ij}$ takes values in the interval $(0,1)$. Also, consider a regression model with the following structure:

\begin{equation}\label{beta1} G(\mbox{E}(\textbf{y}_i|\textbf{u}_i)) = \textbf{x}_i \bm{\beta} + \textbf{z}_i \textbf{u}_i, \end{equation}
for $i = 1, \dots, n$, where $G(.)$ is a vector function linking the conditional mean response vector $\mbox{E}(\textbf{y}_i|\textbf{u}_i)$ with the linear mixed model $\eta_i = \textbf{x}_i \bm{\beta} + \textbf{z}_i \textbf{u}_i$. In this formulation, $\textbf{x}_i$ is the design matrix of dimension $n_i \times p$ corresponding to the vector $\bm{\beta} = (\beta_1, \dots, \beta_p)^\top$ of regression coefficients (the fixed effects), and $\textbf{z}_i$ is the design matrix of dimension $n_i \times q$ associated with the vector $\bm{u}_i = (u_{i1}, \dots, u_{iq})^\top$ (the random effects), with distribution $\bm{u}_i \sim N(\bm{0}, \sigma_{u}^{2} \textbf{I})$. More details about the beta regression model and its extensions with random effects can be found in \citep{mcculloch2004generalized}, \citep{dunn2018generalized}, \citep{figueroa2013mixed}, \citep{verkuilen2012mixed}, \citep{galvis2014augmented}, \citep{grun2012extended}, and \citep{cribari2010beta}.

\subsection{Specific model for an attribute} \label{method2}

To establish the notation, consider a sensory design in balanced incomplete blocks, containing $v$ varieties or treatments arranged in $b$ blocks (panelists), in which each block evaluates $k$ varieties, which are repeated $r$ times, such that $bk = rv$ and $\lambda(v - 1) = r(k - 1)$, where $\lambda$ defines the number of times a pair of varieties appears in the same block. Also, suppose that $L$ sensory attributes are to be evaluated using a hedonic scale with $c$ points; generally, $c = 9$.

To use the beta regression model, the ratings assessed on the $c$-point hedonic scale must be converted to the unit interval $(0, 1)$. Given that the model is specified to handle data where the response variable falls within a bounded open interval, the equation proposed by \citep{smithson2006better}, expressed as $(y(n - 1) + 0.5) / n$, was employed to address the extreme values of 1, where $n = bk$ represents the number of observations.

In this context, as a particular case of equation (\ref{beta1}), the specific beta regression model for an attribute $L$ is given by:

\begin{equation} \label{beta2} \eta_{i} = \ln \left[\frac{\mu_{i}}{1 - \mu_{i}}\right] = (\alpha_{0} + \textbf{u}_{i}) + \bm{x}_{i}^{\top} \bm{\beta}, \end{equation}

where

\begin{equation} \label{modseparado} \mu_{i} = \frac{\exp(\eta_{i})}{1 + \exp(\eta_{i})} = \frac{\exp[(\alpha_{0} + \textbf{u}_{i}) + \bm{x}_{i}^{\top} \bm{\beta}]}{1 + \exp[(\alpha_{0} + \textbf{u}_{i}) + \bm{x}_{i}^{\top} \bm{\beta}]}, \end{equation}
is the mean conditional on the random effect $\textbf{u}_{i} \sim \mbox{N}(\textbf{0}, \sigma_{u}^{2} \textbf{I})$; $\textbf{x}_{i} = (x_{it_{1}}, \dots, x_{it_{k}})^{\top}$, with $1 \leq t_{1}, t_{2}, \ldots, t_{k} \leq v$, corresponding to the variety (treatment) index in each block (panelist).

Thus, in the individual analysis for a given attribute, the following dimensions are considered: $\mathbf{X}(bk \times v)$, and the vectors $\mathbf{y}(bk \times 1)$ and $\bm{\beta}~(v \times 1)$, for the model matrices. The columns of the design matrix $(\mathbf{X})$ contain dummy variables for the varieties (treatments, formulations).

\subsection{Unified Beta Regression Model Approach} \label{unifi}

In this section, we present our proposed method to unify the analysis of all sensory attributes. The central idea of this procedure is based on building a stacked data structure with the inclusion of indicator covariates for the attributes.
For this purpose, let
$$\textbf{Y}= (\textbf{y}_{1}, \textbf{y}_{2}, \dots, \textbf{y}_{L})^{T}$$ 
be the vector that represents the response variable for all attributes, with dimension $(bkL \times 1)$, corresponding to all vectors [$\mathbf{y}~(bk \times 1)$] of the individual analysis stacked. And:
   \begin{equation} \label{matrizes}
\textbf{X}^* = \left[
\begin{array}{c|c}
\textbf{X}_1 & \textbf{X}_2 
\end{array}
\right]
\end{equation}
corresponds to the complete matrix of the unified model, with dimension $(bkL \times v + L - 1)$, in which $\textbf{X}_{1}~(bkL \times v)$ corresponds to the stacked matrices of the individual cases, and $\textbf{X}_{2}~(bkL \times L - 1)$ is the portion of the matrix composed of dummy variables referencing the attributes. With the inclusion of random effects, the linear predictor of the unified mixed beta model, in matrix terms, is given by:

\begin{equation}\label{unifmodel}
 \bm{\eta}=\textbf{X}^{*}\bm{\theta} +\textbf{Z}_{i}\textbf{u}_{i},   
\end{equation}
in which $\bm{\theta}=(\bm{\beta},\bm{\delta})^{T}$ is the parameter vector associated with the components of $\textbf{X}_1$ and $\textbf{X}_2$, respectively, that is, the coefficients for the formulations and attributes. The matrix $\textbf{Z}_{i}~(bkL \times b)$ refers to the random effects, and $\textbf{u}_{i}(b \times 1)$ are distributed as $N(\textbf{0}, \sigma_{u}^{2}\textbf{I})$. The vector $\bm{\eta}(bkL \times 1)$ represents the linear predictor.

The likelihood function of the mixed beta regression model (equation \ref{unifmodel}) is described as follows. The contribution to the likelihood function from individual $i$ is given by:

\begin{equation}
    f_{i}(\mathbf{y}_{i}|\bm{\theta} ,\sigma_{u}^{2}, \phi )=\int \prod_{j=1}^{kL}f_{ij}(y_{ij}|\textbf{u}_{i},\bm{\theta}, \phi )f(\textbf{u}_{i}|\sigma_{u}^{2})\textit{d}\mathbf{u}_{i}, 
    \end{equation}
from which the likelihood for $\bm{\theta}, \sigma_{u}^{2} $ and $\phi$ is derived as: 

\begin{equation}\label{max}
L(\bm{\theta},  \sigma_u^2, \phi) = \prod_{i=1}^{b}f_{i}(\mathbf{y}_{i}|\bm{\theta} ,\sigma_{u}^{2}, \phi )=\prod_{i=1}^{b} \int \prod_{j=1}^{kL}  f_{ij}(y_{ij} \mid  \bm{\theta}, \textbf{u}_i,\phi)  f(\textbf{u}_i \mid \sigma_u^2) \, d\textbf{u}_i.
\end{equation}

Apply logarithm in equation (\ref{max}):

\begin{equation} \label{logver}
l(\bm{\theta}, \sigma_u^2, \phi) = \sum_{i=1}^{b} \ln \left[ \int \prod_{j=1}^{kL} f_{ij}(y_{ij} \mid \bm{\theta}, \mathbf{u}_i, \phi) f(\mathbf{u}_i \mid \sigma_u^2) \, d\mathbf{u}_i \right],
\end{equation}
where $f_{ij}(y_{ij} \mid \bm{\theta}, \textbf{u}_i, \phi)$ represents the conditional density function of the observed data, which depends on the fixed effects $\bm{\theta}$ and the parameter $\phi$, and $f(\textbf{u}_i \mid \sigma_u^2)$ is the distribution of the random effects $\textbf{u}_i$.

The density function of the beta distribution for an observation $y_{ij}$ is given by:
\begin{equation} \label{densidade1}
    f(y_{ij} \mid \mu_{ij}, \phi) = 
    \frac{\Gamma(\phi)}{\Gamma(\mu_{ij}\phi)\Gamma((1-\mu_{ij})\phi)} 
    y_{ij}^{\mu_{ij} \phi - 1} (1 - y_{ij})^{(1 - \mu_{ij}) \phi - 1}
\end{equation}
in which, $\mu_{ij} = \frac{\exp(\eta_{ij})}{1 + \exp(\eta_{ij})}$ is the mean of the beta distribution;
$\eta_{ij} = \mathbf{X}_{ij}^{*\top} \bm{\theta} + \mathbf{Z}_{ij}^{\top} \mathbf{u}_i$ is the linear predictor. Since $\textbf{u}_i \sim N(\textbf{0}; \sigma_{u}^{2}\textbf{I})$ the density function of the random effect is defined by:
\begin{equation} \label{densidade2}
f(\textbf{u}_i \mid \sigma_u^2) = \frac{1}{(2\pi \sigma_u^2)^{q/2}} \exp \left( -\frac{\textbf{u}_i^T\textbf{u}_i}{2 \sigma_u^2} \right).
\end{equation}

Substituting equations (\ref{densidade1}) and (\ref{densidade2}) into equation (\ref{logver}), we obtain:
\begin{align} \label{logvero}
l(\bm{\theta}, \sigma_u^2, \phi) = \sum_{i=1}^{b} \ln \Bigg[ 
\int \Bigg\{ \prod_{j=1}^{kL} 
\frac{\Gamma(\phi)}{\Gamma(\mu_{ij}\phi)\Gamma((1 - \mu_{ij})\phi)} 
y_{ij}^{\mu_{ij}\phi - 1}(1 - y_{ij})^{(1 - \mu_{ij})\phi - 1} \Bigg\} \notag \\
\times 
\Bigg\{\frac{1}{(2\pi \sigma_u^2)^{q/2}} 
\exp\left( -\frac{1}{2\sigma_u^2} \mathbf{u}_i^\top \mathbf{u}_i \right)
\, d\mathbf{u}_i\Bigg\} \Bigg]
\end{align}

The central challenge in maximizing equation (\ref{logvero}) lies in the need to compute $b$ integrals involving the random effects $\bm{u}_{i}$. These integrals do not have a closed-form solution, which is a key step in the estimation process \citep{molenberghs2005models}. The maximum likelihood method was used to estimate the parameters of the beta regression model with random effects (equation \ref{unifmodel}). This method is widely recognized and is the most popular for estimating unknown regression parameters in the beta regression model \citep{abonazel2022dawoud}. To obtain more accurate approximations of the maximum likelihood estimates of the model parameters, the Laplace approximation was employed, implemented in the \texttt{glmmTMB} package, as described by \citep{magnusson2017package, brooks2017glmmtmb}. For more details on the \texttt{TMB} package, which performs the model estimates, refer to \citep{kristensen2015tmb}. All mentioned packages are available in the \texttt{R} software \citep{r2013r}.

The Laplace approximation essentially uses a Taylor expansion to approximate the Gaussian integral over $\textbf{u}_{i}$ around the value that maximizes the likelihood function \citep{joe2008accuracy}. It is an effective method for solving integrals involving a density function with an exponential term and a normal distribution. According to \citep{molenberghs2005models}, Laplace's method was designed to approximate integrals of the form:
\begin{equation}
I = \int e^{Q(\textbf{u})} \, d\textbf{u}
\label{eq:laplace_integral}
\end{equation}
where $Q(\textbf{u})$ is a known, unimodal, and bounded function of a $q$-dimensional variable $\textbf{u}$. Let $\hat{\textbf{u}}$ be the value of $\textbf{u}$ at which $Q$ is maximized. The second-order Taylor expansion of $Q(\textbf{u})$ is then given by:

\begin{equation}
Q(\textbf{u}) \approx Q(\hat{\textbf{u}}) + \frac{1}{2}(\textbf{u} - \hat{\textbf{u}})^\top Q''(\hat{\textbf{u}})(\textbf{u} - \hat{\textbf{u}})
\label{eq:taylor_expansion}
\end{equation}
where $Q''(\hat{\textbf{u}})$ is the Hessian matrix of $Q$, i.e., the matrix of second-order partial derivatives of $Q$ evaluated at $\hat{\textbf{u}}$.

Replacing $Q(\textbf{u})$ in \eqref{eq:laplace_integral} with its approximation from \eqref{eq:taylor_expansion}, we obtain the Laplace approximation:

\begin{equation}
I \approx (2\pi)^{q/2} \left| -Q''(\hat{\textbf{u}}) \right|^{-1/2} e^{Q(\hat{\textbf{u}})}
\label{eq:laplace_approx}
\end{equation}

Each integral in (\ref{logvero}) is proportional to an integral of the form \eqref{eq:laplace_integral}, with the function $Q(\textbf{u})$ defined by:

\[
Q(\mathbf{u}) = \sum_{j=1}^{kL} \log f(y_{ij} \mid \mu_{ij}, \phi) - \frac{1}{2\sigma_u^2} \mathbf{u}^\top \mathbf{u}
\]

Laplace’s method can be used here because the integrand has the required form. Since the mode $\hat{\textbf{u}}$ depends on the unknown parameters $\bm{\theta}$, $\phi$, and $\sigma_u^2$, it must be re-estimated at each step of the likelihood maximization. The approximation becomes exact if $Q(\textbf{u})$ is quadratic, that is, when the integrand is a Gaussian kernel.

\section{Simulation Studies}\label{simulation}

To evaluate the effectiveness of the unified beta regression model in the analysis of sensory data, a detailed simulation study was conducted. The main objective was to examine the agreement between the unified model and the separate models across several experimental scenarios. To optimize the simulation process, three distinct formulations were employed, along with two sensory attributes organized in a balanced complete block design.

Simulated data were generated based on parameters from the mixed cumulative logit model with proportional odds. This choice enabled the assessment of the accuracy of the unified beta model in handling typical data structures found in sensory experiments. The parameterization included random effects to account for heterogeneity across blocks and adjusted conditional probabilities to represent the varying levels of sensory attributes. The cumulative proportional odds model is given by:
\begin{equation}\label{modelologito}
    \eta _{i}= ln \left [ \frac{\gamma  _{j}}{1-\gamma _{j}} \right ]=\alpha _{j}+\bm{\beta}^{T}\mathbf{X}+ \bm{\delta}^{T}\bm{Z}+\mathbf{u}_{i},
\end{equation}
with $\alpha_{j}$ being the intercept of the $j$-th response category referring to a given sensory attribute (acidity, color, aroma, sweetness, and flavor), $\bm{\beta}$ and $\bm{\delta}$ the regression parameter vectors associated with the fixed-effects design matrices $\mathbf{X}$ (formulations) and $\mathbf{Z}$ (attributes), respectively, and $\mathbf{u}_{i}$ the random effect associated with the $i$-th panelist, where $\mathbf{u}_{i} \sim N(\textbf{0}, \sigma_{u}^{2} \textbf{I})$. The reference category for the response was "1 = extremely dislike", with attribute $A$ evaluated for formulation $F_{1}$.

Thirteen distinct scenarios were considered, with 1,000 data replications for each scenario, and two different sample sizes, $N = 90$ and $N = 300$ panelists (blocks). One specific scenario, characterized by $(f_{1} < f_{3} < f_{2})$, included the following parameter settings:

\begin{equation*}
\begin{aligned}
\bm{\theta}_{j} &= (\alpha_{j}, \beta_{[f3]}, \beta_{[f2]}, \gamma_{[A]}) \\
\bm{\theta}_{2} &= (5, 0, 0, 0.5) \\
\bm{\theta}_{3} &= (0, 0, 0, 1) \\
\bm{\theta}_{4} &= (0, 6, 0, 0.5) \\
\bm{\theta}_{5} &= (0, 0, 7, 0.5)
\end{aligned}
\end{equation*}

The concordance rates between the unified model (Equation \ref{unifmodel}) and the beta regression models with random effects, evaluated separately for each sensory attribute (Models A and B as defined in Equation (\ref{beta2})), are presented in Table \ref{tabelasimulation}.

\begin{table}[H]
\centering
\caption{Concordance rates between the unified mixed beta regression model (Equation \ref{unifmodel}) and the mixed beta regression model, evaluated for each sensory attribute separately (Model for sensory attribute A and Model for sensory attribute B (Equation \ref{beta2})) across the thirteen different simulation scenarios, considering the three formulations $(F_{1}, F_{2}, F_{3})$ for sample sizes $N=90$ and $N=300$.}
\label{tabelasimulation}
\begin{tabular}{lccc}
\toprule
\multicolumn{4}{c}{$N = 90$} \\
\cmidrule{2-4}
Scenarios & Unified Model & Model for Attribute A & Model for Attribute B \\
\midrule
$F_{3}<F_{1}<F_{2}$ & 100.0\% & 99.9\% & 100.0\% \\
$F_{1}<F_{3}<F_{2}$ & 100.0\% & 100.0\% & 100.0\% \\
$F_{2}=F_{3}<F_{1}$ & 99.9\% & 100.0\% & 99.5\% \\
$F_{2}<F_{3}<F_{1}$ & 100.0\% & 100.0\% & 99.9\% \\
$F_{3}<F_{2}<F_{1}$ & 100.0\% & 100.0\% & 99.8\% \\
$F_{1}=F_{2}=F_{3}$ & 97.4\% & 99.7\% & 97.6\% \\
$F_{1}=F_{2}<F_{3}$ & 99.8\% & 99.8\% & 100.0\% \\
$F_{1}<F_{2}=F_{3}$ & 92.4\% & 92.2\% & 92.7\% \\
$F_{1}<F_{2}<F_{3}$ & 100.0\% & 99.9\% & 99.9\% \\
$F_{2}<F_{1}=F_{3}$ & 92.7\% & 91.7\% & 91.6\% \\
$F_{2}<F_{1}<F_{3}$ & 100.0\% & 100.0\% & 100.0\% \\
$F_{3}=F_{1}<F_{2}$ & 99.6\% & 100.0\% & 99.8\% \\
$F_{3}<F_{1}=F_{2}$ & 92.7\% & 90.8\% & 93.2\% \\
\midrule
\multicolumn{4}{c}{$N = 300$} \\
\cmidrule{2-4}
$F_{3}<F_{1}<F_{2}$ & 100.0\% & 100.0\% & 100.0\% \\
$F_{1}<F_{3}<F_{2}$ & 100.0\% & 100.0\% & 100.0\% \\
$F_{2}=F_{3}<F_{1}$ & 97.5\% & 99.6\% & 98.3\% \\
$F_{2}<F_{3}<F_{1}$ & 100.0\% & 100.0\% & 100.0\% \\
$F_{3}<F_{2}<F_{1}$ & 100.0\% & 100.0\% & 100.0\% \\
$F_{1}=F_{2}=F_{3}$ & 82.5\% & 96.7\% & 88.3\% \\
$F_{1}=F_{2}<F_{3}$ & 99.1\% & 99.7\% & 99.2\% \\
$F_{1}<F_{2}=F_{3}$ & 93.6\% & 91.5\% & 92.4\% \\
$F_{1}<F_{2}<F_{3}$ & 100.0\% & 100.0\% & 100.0\% \\
$F_{2}<F_{1}=F_{3}$ & 92.6\% & 92.3\% & 92.4\% \\
$F_{2}<F_{1}<F_{3}$ & 100.0\% & 100.0\% & 100.0\% \\
$F_{3}=F_{1}<F_{2}$ & 98.4\% & 99.8\% & 98.4\% \\
$F_{3}<F_{1}=F_{2}$ & 93.2\% & 92.1\% & 93.7\% \\
\bottomrule
\end{tabular}
\end{table}

TTable \ref{tabelasimulation} presents the simulation results obtained from hypothetical scenarios involving hierarchical relationships among the formulations $F_{1}$, $F_{2}$, and $F_{3}$. The formulations were ranked based on assumed quality, generating scenarios such as $F_{3}<F_{1}<F_{2}$, where $F_{2}$ is considered to have the highest quality, followed by $F_{1}$, and finally $F_{3}$. For each scenario, concordance between the unified model and the separate models was calculated, considering both the individual and joint evaluations of the sensory attributes.

The table demonstrates high concordance rates, exceeding 90\% in most scenarios, indicating strong agreement between the unified model and the models evaluated separately by sensory attribute. Simpler scenarios, or those in which the formulations ($F_1$, $F_2$, $F_3$) exhibit well-defined hierarchies, yielded the best results. For example, in the scenario $F_{1}<F_{3}<F_{2}$, where $F_{1}$ is inferior to the other formulations, the observed concordance rates were consistently 100.0\% across all simulated models and panelist sample sizes.

When the sample size was increased to $N = 300$, the concordance rates did not show significant improvement, indicating that expanding the number of evaluators does not substantially affect the agreement between models. This finding suggests that the unified model is efficient in capturing variation between formulations, even in scenarios involving multiple sensory attributes, reinforcing its applicability and robustness for studies of this nature.

Furthermore, the consistency of concordance rates across different sample sizes highlights the stability of the unified model, making it a valuable tool for sensory analyses that require high precision and integration of multiple attributes.

All simulations were conducted using the R software \citep{core2019r}, with the \texttt{glmmTMB} package \citep{magnusson2017package, brooks2017glmmtmb} employed to fit the beta regression models.

\section{Applications and Results} \label{aplication}

The dataset used in this study comes from an experiment conducted in 2025 at the Sensory Analysis Laboratory of the Department of Agroindustry, Food, and Nutrition, “Luiz de Queiroz” College of Agriculture (ESALQ/USP), Piracicaba, Brazil. Eight brands of grape juice were evaluated, equally divided into whole juices (100\% juice or not-from-concentrate) and non-whole juices (from concentrate or reconstituted).

For the exploratory analyses, bar charts and boxplots were constructed to visually represent the data. The ratings assigned by each panelist were based on a five-point hedonic scale, ranging from "1 = extremely disliked or disliked very much" to "5 = extremely liked or liked very much." The beverage formulations (not-from-concentrate juices: $F_{873}$, $F_{661}$, $F_{419}$, $F_{571}$; reconstituted juices: $F_{715}$, $F_{179}$, $F_{732}$, $F_{318}$) were evaluated across five sensory attributes: (1) color, (2) acidity, (3) sweetness, (4) flavor, and (5) aroma.

\begin{figure}[H]
  \centering
  \begin{adjustbox}{width=1\textwidth,center}
    \includegraphics{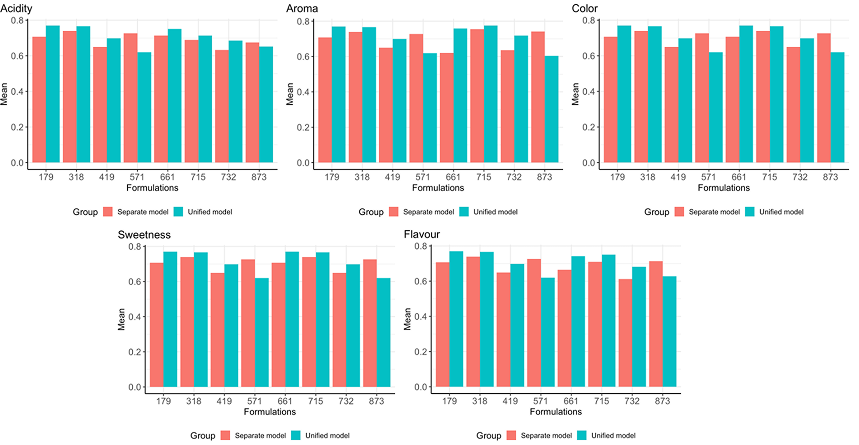}
  \end{adjustbox}
  \caption{Bar chart showing the means of the stacked data for the unified model and the data separated by sensory attribute for the eight grape juice formulations, based on the data from the study conducted at ESALQ/USP in 2025.}
  \label{barras}
\end{figure}

A higher average for all evaluated sensory attributes was identified in formulations $F_{179}$, $F_{715}$, and $F_{318}$, as illustrated in Figure \ref{barras}. The unified model, based on the stacked data, reinforces these findings, suggesting greater sensory acceptability of these formulations compared to the others evaluated.

\begin{figure}[H]
  \centering
  \begin{adjustbox}{width=0.90\textwidth,center}
  \includegraphics{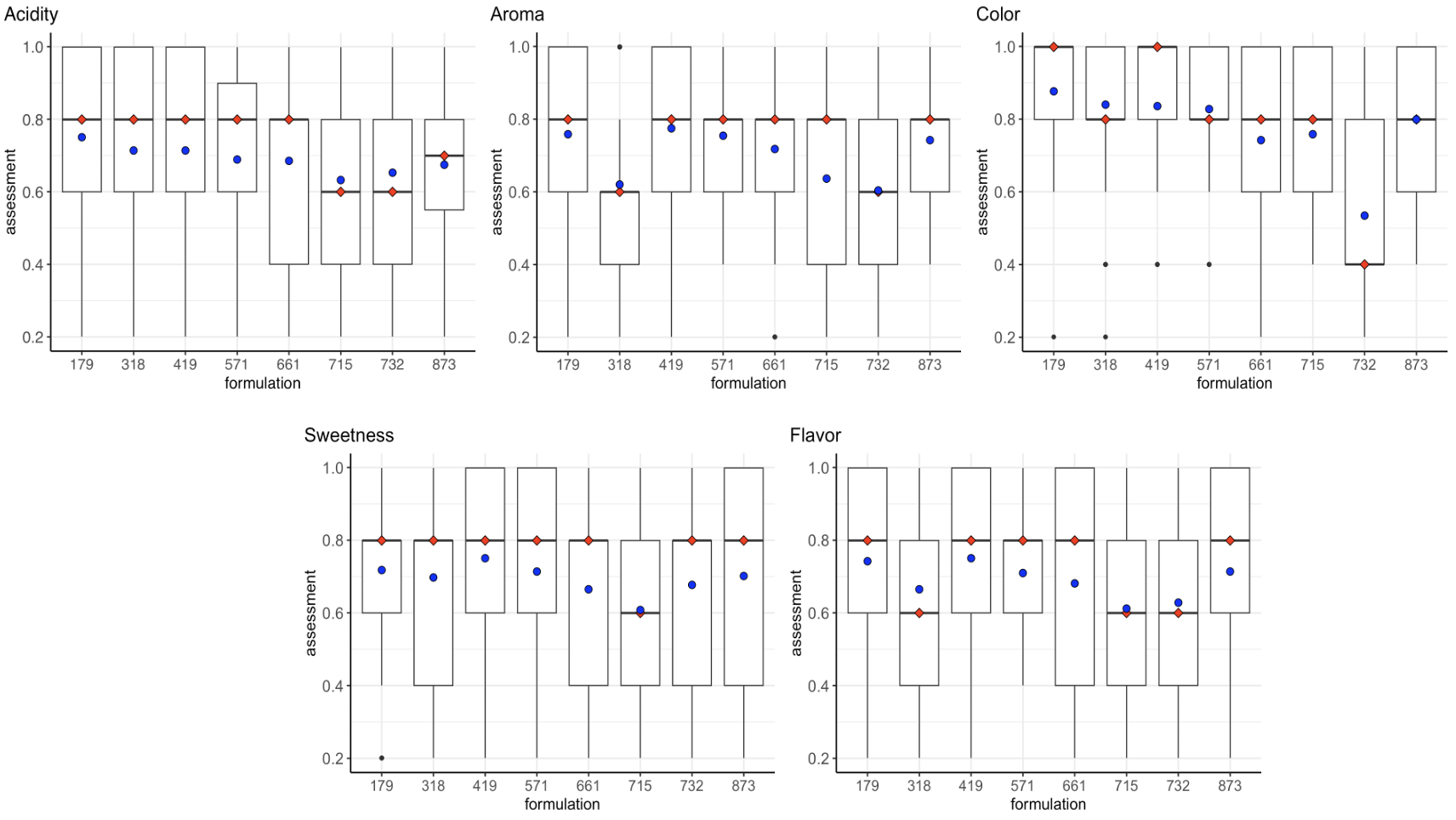}
  \end{adjustbox}
  \caption{Boxplots of the sensory attributes for the eight grape juice formulations derived from the data collected at ESALQ/USP in 2025.}
  \label{figura3}
\end{figure}

The boxplots in Figure~\ref{figura3} display the mean and median values for each formulation and sensory attribute, represented by blue and red points, respectively. It is observed that, in several cases, the medians are equal to or greater than the means, suggesting possible asymmetry in the data distribution. The highest means among the evaluated attributes were observed in formulations $F_{179}$, $F_{419}$, and $F_{571}$. Additionally, outliers were identified in the aroma, color, and sweetness attributes, indicating the presence of atypical values that may influence the analysis.

\begin{figure}[H]
\centering \hspace{-2em}
\begin{adjustbox}{width=0.40\textwidth,center}
  \includegraphics{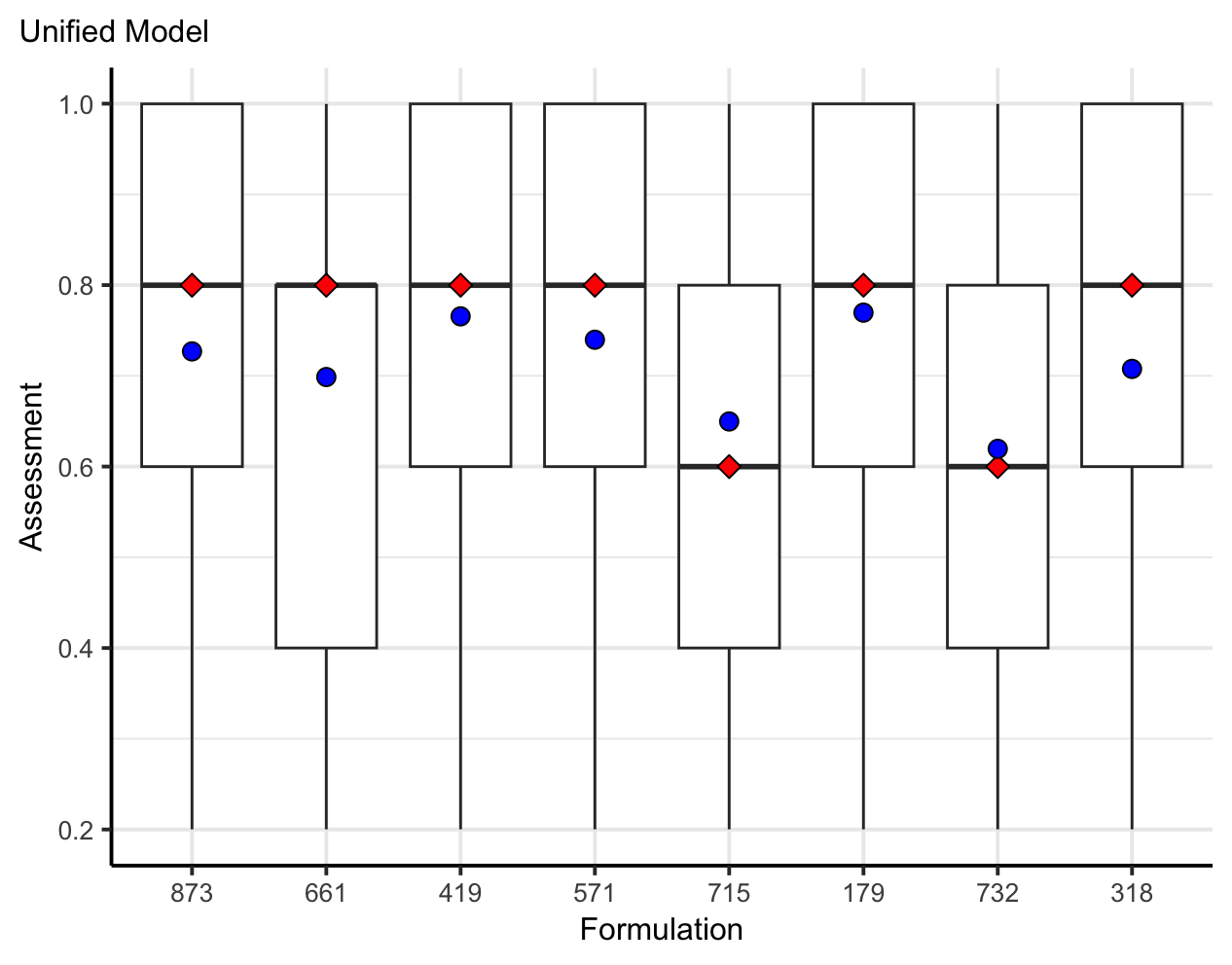}
  \end{adjustbox}
  \caption{Boxplot of the unified model for the eight grape juice formulations, derived from data collected at ESALQ/USP in 2025.} 
  \label{figura4}
\end{figure}

As depicted in Figure~\ref{figura4}, the asymmetry of the data and the presence of outliers are evident. Formulations $F_{419}$, $F_{179}$, and $F_{571}$ stand out by exhibiting the highest mean values.

These analyses, however, are exploratory in nature. Subsequently, as described in Section~\ref{method}, beta regression models were fitted to the data for all sensory attributes, both individually and through a unified model. The results of the individual model fits are omitted.

For the adjustment of models using the stacked data, that is, in the unified model, the Null Model: $\eta_{0} = \alpha_{0}$, which included only the intercept, was initially considered. Subsequently, Model 1: $\eta_{1} = \alpha_{0} + \bm{X}\bm{\beta}$ incorporated the fixed effect of formulation, where $\bm{X}$ denotes the formulation covariates and $\bm{\beta}^{T} = (\beta_{1}, \ldots, \beta_{v})$ is the associated parameter vector. Model 2: $\eta_{2} = \alpha_{0} + \bm{X}^{*}\bm{\theta}$, where $\bm{\theta} = (\bm{\beta}, \bm{\delta})^{T}$, with $\bm{\delta}^{T} = (\delta_{1}, \ldots, \delta_{L})$ denoting the vector of parameters associated with the attribute covariates, and $\bm{X}^{*}$ as defined in the matrix expression~\ref{matrizes}. Finally, Model 3, defined by Equation~(\ref{unifmodel}), the most comprehensive, included both fixed effects and the random effect of the panelists. The selection of the best model structure was based on the estimated dispersion parameter ($\hat{\phi}$), the log-likelihood function (log(L)), and the AIC, as presented in Table~\ref{tabela2}.

\begin{table}[H]
\centering
\caption{Dispersion parameter ($\hat{\phi}$), maximized log-likelihood (log(L)), and Akaike Information Criterion (AIC) for the unified model, considering four model structures: Null, Model 1, Model 2, and Model 3, based on the study conducted at ESALQ/USP in 2025.}
\vspace{0.2cm}
\begin{tabularx}{\textwidth}{l p{6cm} c c c}
\hline
 & Models & $\hat{\phi}$ & $\log(L)$ & AIC \\ \hline
 & Null~(intercept) & 1.44  & 1676.8 & -3349.7 \\
 & $1$~(intercept + formulation effect) & 1.51 & 1719.1 &  -3420.2  \\
 & $2$~(intercept + formulation effect + attribute effect) & 1.54 & 1743.6 & -3461.1 \\
 & $3$~(intercept + formulation effect + attribute effect + random effects) & 1.77 &  1808.4  & -3588.8\\ \hline
\end{tabularx}
\label{tabela2}
\end{table}

An increase in both the log-likelihood and the dispersion parameter was observed as model complexity increased, with the corresponding values presented in Table~\ref{tabela2}. The inclusion of random effects for the panelists proved important in the context of the BIB design, as evidenced by the lower AIC values. The likelihood ratio test between the models (Model 2 vs. Model 3) resulted in a \textit{p}-value less than 0{.}01, confirming that the functional structure of the data was better represented by the mixed model. Furthermore, the formulation effect was found to be statistically significant (\textit{p}-value $< 0.05$). Consistent results were obtained in the models fitted separately for each sensory attribute; however, for the sake of conciseness, these details were not included in this paper.

Based on these findings, the unified mixed effects model was selected. The corresponding parameter estimates, standard errors, and $p$-values are provided in Table~\ref{parameters}. The second column of Table~\ref{parameters} presents the estimated values of the parameters $\hat{\beta}_{v}$ and $\hat{\delta}_{L}$, corresponding to the effects of the formulations ($v=1,2,\ldots,8$) and the sensory attributes ($L=1,2,\ldots,5$), respectively. Formulation $F_{179}$ was used as the reference category for the formulation factor, while level (2) was adopted as the reference for the attribute factor. The highest estimated values of $\hat{\beta}_{v}$ were associated with the formulations receiving the greatest acceptance, such as $F_{419}$, $F_{179}$, and $F_{873}$. In contrast, the lowest values were attributed to formulations with the least favorable sensory evaluations. Based on the estimated coefficients, it becomes possible to calculate the probabilities associated with each combination of formulation and attribute. For example, to estimate $P(\beta=F_{419}, \delta=1)$, the values $\hat{\alpha}_{0} = 1{.}36$, $\hat{\beta}_{419} = 0{.}06$, and $\hat{\delta}_{1} = 0{.}47$ were used, in addition to the random effect, as described in Equation~(\ref{unifmodel}). Similarly, the probabilities corresponding to the other formulations and attributes can be obtained.

\begin{table}[H]
\caption{Estimated parameters, standard errors (S.E.), and $p$-values from the unified mixed beta regression model for the analysis of sensory attributes in the study conducted at ESALQ/USP in 2025.}
\label{parameters}
\centering
\begin{tabular}{lccc}
\toprule
\textbf{Parameter} & \textbf{Estimate} & \textbf{S.E.} & \textbf{$p$-value} \\
\midrule
$\alpha_0$         & 1.36  & 0.10 & 0.00 \\
$\beta_{732}$      & -0.62 & 0.11 & 0.00 \\
$\beta_{318}$      & -0.32 & 0.11 & 0.00 \\
$\beta_{419}$      & 0.06  & 0.11 & 0.54 \\
$\beta_{571}$      & -0.23 & 0.11 & 0.03 \\
$\beta_{661}$      & -0.26 & 0.11 & 0.01 \\
$\beta_{715}$      & -0.72 & 0.11 & 0.00 \\
$\beta_{873}$      & -0.23 & 0.11 & 0.03 \\
$\delta_1$         &  0.47  & 0.08 & 0.00 \\
$\delta_3$         &  0.01  & 0.08 & 0.90 \\
$\delta_4$         &  0.02  & 0.08 & 0.79 \\
$\delta_5$         & -0.02 & 0.08 & 0.76 \\
\bottomrule
\end{tabular}
\end{table}

The estimated values obtained from the model (\ref{unifmodel}) for all formulations and sensory attributes evaluated are presented in Table \ref{valoresestimados}.

\begin{table}[H]
\caption{Estimated means from the unified model (Equation~\ref{unifmodel}) for all beverage formulations and sensory attributes in the study conducted at ESALQ/USP in 2025.}
\label{valoresestimados}
\centering
\begin{tabular}{l|cccccccc}
\toprule
\multirow{2}{*}{Sensory Attributes} & \multicolumn{8}{c}{Beverage formulations} \\
\cmidrule{2-9}
& $F_{732}$ & $F_{318}$ & \cellcolor{gray!10}$F_{419}$ & $F_{571}$ & $F_{661}$ & $F_{715}$ & \cellcolor{gray!10}$F_{873}$ & \cellcolor{gray!10}$F_{179}$ \\
\midrule
Color       & 0.77 & 0.82 & \cellcolor{gray!10}0.87 & 0.83 & 0.82 & 0.75 & \cellcolor{gray!10}0.83 & \cellcolor{gray!10}0.86 \\
Aroma       & 0.67 & 0.73 & \cellcolor{gray!10}0.80 & 0.75 & 0.74 & 0.65 & \cellcolor{gray!10}0.75 & \cellcolor{gray!10}0.79 \\
Sweetness   & 0.68 & 0.74 & \cellcolor{gray!10}0.81 & 0.76 & 0.75 & 0.65 & \cellcolor{gray!10}0.76 & \cellcolor{gray!10}0.80 \\
Flavor      & 0.68 & 0.74 & \cellcolor{gray!10}0.81 & 0.76 & 0.75 & 0.66 & \cellcolor{gray!10}0.76 & \cellcolor{gray!10}0.80 \\
Acidity     & 0.79 & 0.74 & \cellcolor{gray!10}0.80 & 0.75 & 0.75 & 0.65 & \cellcolor{gray!10}0.76 & \cellcolor{gray!10}0.80 \\
\bottomrule
\end{tabular}
\end{table}

According to Table~\ref{valoresestimados}, the highest estimated mean was observed for formulation $F_{419}$, followed by $F_{179}$ and $F_{873}$. In contrast, lower values were recorded for formulations $F_{715}$ and $F_{732}$, indicating lower acceptance or less favorable evaluations of the sensory attributes associated with these formulations.

In general, the sensory attribute color received the highest evaluation, whereas aroma showed the lowest estimated means. These results suggest that consumers are more sensitive to visual aspects, while the product’s aroma has a relatively lower influence on their evaluations. Therefore, improvement strategies could focus on enhancing the aroma attribute to increase overall appeal.

The higher evaluation of formulations $F_{419}$ and $F_{873}$ (100\% juice) reflects greater acceptability among the assessors for formulations with a higher proportion of juice relative to sugar. In contrast, the lowest-rated formulations, $F_{715}$ and $F_{732}$ (non-100\% juice), indicate a tendency for assessors to reject products with higher levels of additives such as sugar, suggesting that consumers prefer juices with fewer added ingredients.

This analysis provides a solid foundation for refining product development strategies and prioritizing the sensory characteristics most valued by consumers. Similar results were observed in the individual evaluation of sensory attributes.

\begin{figure}[H]
  \centering
  \includegraphics[width=0.8\textwidth]{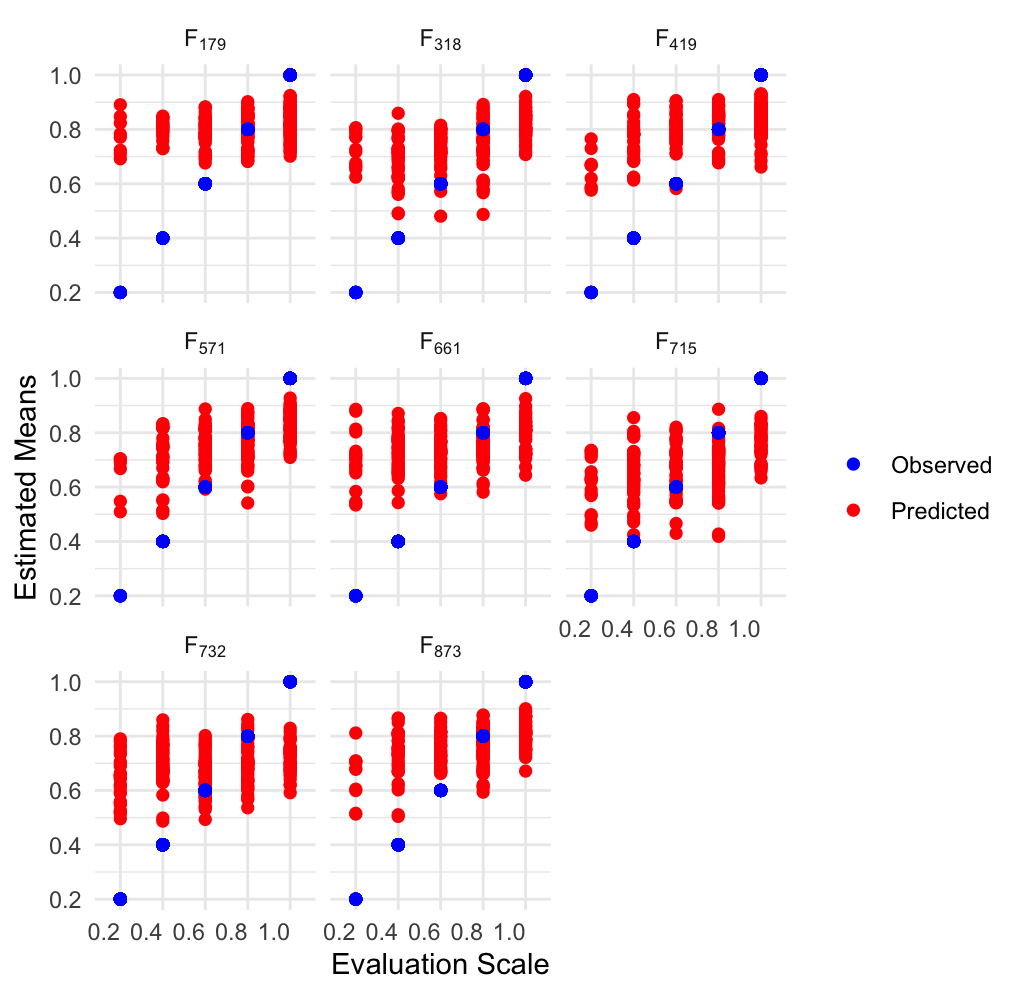}
  \caption{Observed values and values estimated by the beta regression model with random effects for prebiotic beverage formulations, considering five sensory attributes from the study conducted at ESALQ/USP in 2025.}

  \label{dispersao}
\end{figure}

Good agreement between the observed and predicted values can be seen in Figure~\ref{dispersao}, although some discrepancies were noted at the lower end of the scale.

\section{Conclusion} \label{conclusion}

This study proposed a unified beta regression model to enable the simultaneous analysis of multiple sensory attributes within a single structure. This formulation allows for a more compact data representation, avoiding excessive parameterization and reducing the complexity typically associated with separate models for each attribute. The results obtained in the simulation studies demonstrated that this unified strategy was consistent with those of separate analyses by sensory attribute, supporting its suitability for sensory data. Concentrating the information into a single model facilitates interpretation, simplifies comparison between formulations, and streamlines the overall analytical process. Further studies may explore applications in broader experimental scenarios, assess alternative link functions, and develop tools for handling observations at the limits of the scale. These extensions may contribute to expanding the applicability of this modelling approach in sensory evaluation.

\bmhead{Acknowledgments}
This publication has emanated from research conducted with the financial support of the Brazilian foundation Coordenação de Aperfeiçoamento de Pessoal de Nível Superior (CAPES), process number 88887.821274/2023-00, and Taighde Eireann – Research Ireland under Grant Number 18/CRT/6049.

\bibliography{bibliography}

\end{document}